# AN ACTIVE HOST-BASED INTRUSION DETECTION SYSTEM FOR ARP-RELATED ATTACKS AND ITS VERIFICATION[1]

Ferdous A Barbhuiya, Santosh Biswas and Sukumar Nandi

Department of Computer Science and Engineering
Indian Institute of Technology Guwahati, India - 781039
`{ferdous,santosh_biswas,sukumar}@iitg.ernet.in`
`http://www.iitg.ernet.in`

## ABSTRACT

*Spoofing with falsified IP-MAC pair is the first step in most of the LAN based-attacks. Address Resolution Protocol (ARP) is stateless, which is the main cause that makes spoofing possible. Several network level and host level mechanisms have been proposed to detect and mitigate ARP spoofing but each of them has their own drawback. In this paper we propose a Host-based Intrusion Detection system for LAN attacks, which works without any extra constraint like static IP-MAC, modifying ARP etc. The proposed scheme is verified under all possible attack scenarios. The scheme is successfully validated in a test bed with various attack scenarios and the results show the effectiveness of the proposed technique.*

## 1. INTRODUCTION

In Local Area network (LAN), when two hosts want to communicate with each other, they need to know the MAC address of each other. If the communicating host does not know MAC address of the destination, it sends an ARP request to the broadcast domain asking for the MAC address corresponding to the destination host's IP address. The destination host identifies that the ARP request is meant for its IP address and hence, sends back its MAC address in a unicast ARP reply packet. Most of the LAN based attacks are launched by sending falsified IP-MAC pairs to the host being targeted. The victim machine assumes the MAC address in the forged ARP packet as the genuine MAC address associated with the IP. Now when the victim machine wants to communicate with the system having the given IP, it sends all packets to the false MAC address (i.e., to a different host the attacker wants the victim to send). Attacks based on falsified IP-MAC pairs are feasible because host updates its ARP cache without verifying the genuineness of the IP-MAC pair of the source [1]. Also, the hosts cache all the ARP replies sent to them even if they had not sent an explicit ARP request for them. In other words, ARP spoofing is possible because of the stateless nature of the ARP. Various mechanisms have been proposed to detect and mitigate these ARP attacks at both the host-level and network-level. In [2], a literature review on network based IDS for detecting ARP spoofing attacks with their drawbacks have been discussed. The authors also present a new network based IDS to detect such spoofing attacks and highlights how many of the drawbacks are eliminated. The present paper is focused towards development of a host based IDS for ARP spoofing based attacks. So, all discussions regarding existing techniques for ARP attack detection are confined to host-based ones; the techniques can be broadly classified as follows.

---

[1] This journal paper is an extended version (invited) of the conference paper "An Active Host-based Detection Mechanism for ARP-related Attacks" by S Roopa, R Ratti, Neminath H, F.A. Barbhuiya, S Biswas, S Nandi, A Sur and V Ramachandran, Indian Institute of Technology - Guwahati, India presented in The Second International Conference on Network & Communications Security (NCS) 2010. The first part of this journal paper "An Active Host-based Detection Mechanism for ARP-related Attacks" is from the conference version and the second part "Verification" is an extension.

### 1.1 Static ARP [3]

ARP related attacks can be avoided completely by manually as-signing static IP addresses to all the systems in the network [3]. The ARP cache of the host has the static mapping of IP-MAC pairings of all other hosts inside the net-work. Since these entries are immutable, any spoofed packets will be blindly ignored by the kernel. However, this solution is not applicable to large networks because of the problem of scalability and management especially in a dynamic environment.

### 1.2 Stateful ARP [4, 5]

The technique is based on extending the existing the standard ARP protocol by modifying the ARP cache from a stateless to a stateful one. Here, the host has a state ARP cache which holds the states of the previous requests and verifies the replies with it. Two queues requestedQ and respondedQ are used to store the state information of cache. Now the incoming responses are matched from corresponding requestedQ and enters into responded queue till timeout.

A major problem in the Stateful ARP based approach is that Gratuitous request/reply, is not supported. Moreover, the modification of stateless cache to stateful cache requires the extension of the protocol specification thereby making it more complex. ARP is basically designed to keep the protocol simple and so, modification to standard ARP is not desirable.

### 1.3 Cryptographic Solutions [6, 7]

Another solution is to utilize the cryptographic techniques to prevent the ARP attacks. The limitation of such techniques is that each host has to maintain the public key for every other host in network. Also, all the hosts inside the network must be configured to understand the new protocol which requires the upgradation of the network stacks of all the systems involved. Moreover, lots of processing overhead for the signature generation, verification and key management is involved.

### 1.4 Signature based IDS [8]

Signature based IDS like Snort [8] can be used to detect ARP attacks, but the main problem here is the generation of a large number of false alarms. Furthermore, the ability of IDSs to detect all forms of ARP related attacks are limited [9].

### 1.5 Software based Solutions [10-12]

Many software solutions are commercially available such as, ARPWATCH, Arp-Guard, X-ARP etc. These software basically maintain a table with IP-MAC associations and any change in the association is immediately reported to the system administrator. The problem with this approach is, if the first sent packet itself is having a spoofed MAC address then the whole system fails. Further, any genuine change in IP-MAC pair will be discarded (e.g., when notified by Gratuitous request and reply).

### 1.6 Active techniques for detecting ARP attacks

In active detection based techniques, probe packets are sent across the network to get the information of the suspected host for which the IP-MAC pair has been changed. Several active techniques for detecting ARP attacks have been reported; they are briefly discussed below.

In [13], a database of known IP-MAC pairs is maintained and on detection of a change the new pair is actively verified by sending a TCP SYN packet as the probe. Also, any new pair of IP-MAC is first verified by the probing technique before entering it in the database. On receiving the probe, the genuine system will respond with SYN/ACK or RST depending upon whether the port is open or not. This scheme is able to detect ARP spoofing attacks but it violates the network layering architecture.

Another means to confirm the authenticity of the receiving ARP response packet is by crafting a RARP request packet [14], which seeks the IP address corresponding to the given MAC address. By comparing the IP addresses of the responses, MAC cloning might be detected. However, a single MAC address may genuinely correspond to a number of IP addresses, in which case, a lot of false positives could be generated. In a similar scheme proposed in [15], ARP probe packets (instead of RARP request packet) are used for IP-MAC validation of the source host from which an ARP packet is received. Unicast ARP probe packets are sent to the host (identified by the MAC) for verification. If a mismatch occurs in the IP addresses of the probe reply compared to the ARP packet being verified, spoofing is notified. The scheme follows the network layering concepts but fails if the attacker is spoofing some IP with its own MAC address. This is because the unicast ARP probe generated by the IDS will go only to the attacker (identified by the MAC) and it would reply back with the same spoofed IP associated with its MAC.

Hence, from the review it may be stated that an ARP attack prevention/detection scheme needs to have the following features

 – Should not modify the standard ARP or violate layering architecture of network – Should generate minimal extra traffic in the network

 – Should detect a large set of LAN based attacks – Hardware cost
 of the scheme should not be high

   In this paper, we propose an active host based IDS (HIDS) to detect a large set of ARP related attacks namely, malformed packets, response spoofing, request spoofing and denial of service. This technique does not require changes in the standard ARP and does not violate the principles of layering structure. Finally, we present a proof of completeness and correctness of the proposed scheme.

   Rest of the paper is organized as follows. In Section 2 we present the proposed approach. In Section 3 we discuss the test bed and experimental results. In Section 4 we prove the correctness of the proposed approach. Finally we conclude in Section 5.

## 2. PROPOSED SCHEME

In this section we discuss the proposed host-based active intrusion detection scheme for ARP related attacks. The following assumption is made regarding the LAN

 – Non-compromised (i.e., genuine) hosts will send a response to an ARP request within a specific interval $T_{req}$.

### 2.1 Data Tables for the Scheme

The proposed IDS running in a host ensures the genuineness of the IP-MAC pairing (of any ARP packet it receives) by an active verification mechanism. The IDS sends verification messages termed as probe requests upon receiving ARP requests and ARP replies. To assist in the probing and separating the genuine IP-MAC pairs with that of spoofed ones, we maintain some information obtained along with the probe requests, ARP requests and ARP replies in some data tables. The information and the data tables used are enumerated below. Henceforth in the discussion, we use the following short notations: *IPS* - Source IP Address, *IPD* - Destination IP Address, *MACS* - Source MAC Address, *MACD* - Destination MAC Address. Fields of any table would be represented by ⟨TableName⟩$_{(field)}$; e.g., $RQT_{IPS}$ represents the source IP filed of "Request-sent table". Also, ⟨TableName⟩$_{MAX}$ represents the maximum elements in the table at a given time.

1. Every time an ARP request is sent from the host querying some MAC address, an entry is created in the "Request-sent table" (denoted as $RQT$) with the destination IP($RQT_{IPD}$) of the ARP packet. Also the time t when the request was sent is recorded in the table as $RQT_\tau$. Its entries timeout after $T_{req}$ seconds. The value of $T_{req}$ will depend on the ARP request-reply round trip time, which can be fixed after a series of experiments on the network. According to [15], the approximate ARP request-reply round trip time in a LAN is about 1.2 ms - 4.8 ms.

2. Every time an ARP reply packet is received by the host from any other system in the network, an entry is created in the "Response-received table" (denoted as $RST$) with its source IP ($RST_{IPS}$) and source MAC ($RST_{MACS}$). Also the time when the response was received is recorded in the table. Its entries timeout after $T_{resp}$ seconds. The $T_{resp}$ value can be determined based on the ARP cache timeout value of the host.

3. When some IP-MAC pair is to be verified, an ARP probe is sent and response is verified. The probe is initiated by the HIDS, upon receiving either a Request or a Response. The source IP address and the source MAC address from the Re-quest/ Response packets used for verification are stored in "Verification table" (de-noted as $VRFT$). The entries in this table are source IP ($VRFT_{IPS}$) and source MAC ($VRFT_{MACS}$).

4. Every time any IP-MAC pair is verified and found to be correct, an entry is created for the pair in the "Authenticated bindings table" (denoted as $AUTHT$). There are two fields in this table, IP address ($AUTHT_{IP}$) and MAC address ($AUTHT_{MAC}$).

## 2.2 Algorithms of the IDS

The proposed HIDS algorithm has two main modules namely, ARP REQUEST-HANDLER() and ARP RESPONSE-HANDLER() to handle incoming and outgoing ARP packets respectively. The modules are discussed below:

Algorithm 1 processes all the ARP request packets received by the host. For any ARP request packet $RQP$ received, the HIDS first checks if it is malformed (i.e., is there any change in the immutable fields of the ARP packer header or different MAC addresses in the MAC and ARP header field) or unicast; if so, a status flag is set accordingly and stops further processing of this packet.

The HIDS next finds whether the packet received is a Gratuitous ARP request and the status flag is set accordingly. Gratuitous ARP request can be determined if $RQP_{IPS} == RQP_{IPD}$. For such Gratuitous ARP request, ARP probe is sent for checking the correctness of the IP-MAC pair. Hence, the VERIFY IP-MAC() module is called for $RQP$ along with t (the time information when $RQP$ was received).

If neither of the above cases match, then $RQP_{IPS}$ is searched in the Authenticated bindings table. If a match is found as $AUTHT_{IPS}[i]$ (where $i$ is the $i^{th}$ entry in the $AUTHT$) and the corresponding MAC address $AUTHT_{MACS}[i]$ in the table is same as $RQP_{MACS}$, the packet has genuine IP-MAC pair which is already recorded in the Authenticated bindings table. In case of a mismatch in the MAC address (i.e., $RQP_{MACS} \neq AUTHT_{MACS}[i]$) the packet is spoofed with a wrong MAC address and hence the status flag is set as spoofed. It may be noted that this checking of spoofing could be done without ARP probe thereby reducing ARP traffic for verification.

Also, it may be the case that IP-MAC pair given in $RQP_{IPS}$ is not verified as yet and no entry can be found in Authenticated bindings table corresponding to $RQP_{IPS}$. In such a case, an ARP probe is to be sent by the HIDS to $RQP_{IPS}$ and $RQP_{MACS}$ for verifying the correctness of the $RQP_{IPS}$-$RQP_{MACS}$ pair. This is handled by the VERIFY IP-MAC() module with $RQP$ and τ as parameters.

### 2.2.1 Algorithm 1: ARP REQUEST HANDLER

**Input:** *RQP* - ARP request packet, t - time at which *RQP* was received, Request-sent table, Verification table, authenticated bindings table

**Output:** Updated Request-sent table, Status

```
 1: if (RQP is malformed) then
 2:     Status=malformed
 3: else if (RQP is Unicast) then
 4:     Status=Unicast
 5: else
 6:    if (RQP_IPS == RQP_IPD) then
 7:        Status=Gratuitous Packet
 8:        VERIFY IP-MAC(RQP, τ)
 9:    else
10:      if (RQP_IPS == AUTHT_IPS[i] (for some i, 1 ≤ i ≤ AUTHT_MAX ) then
11:         if (RQP_MACS == AUTHT_MACS[i]) then
12:            Status= Genuine
13:         else
14:            Status=Spoofed
15:         end if
16:      else
17:         VERIFY IP-MAC(RQP, τ)
18:      end if
19:    end if
20: end if
```

Algorithm 2 is an ARP response handler. For every ARP response packet *RSP* received by the host, the HIDS determines whether the reply is malformed; if malformed, a status flag is set accordingly and the next packet is processed. Otherwise, the source IP ($RSP_{IPS}$), source MAC ($RSP_{MACS}$), and timestamp $\tau$ of the received packet are recorded in the Response-received table. Next, it verifies whether the packet is a Gratuitous ARP reply by checking if $RSP_{IPS} == RSP_{IPD}$. For such a Gratuitous ARP reply, an ARP probe is sent to check the correctness of the IP-MAC pair. Hence, the VERIFY IP-MAC() module is called.

If the reply packet is not Gratuitous, next it verifies if it is a reply for any ARP probe sent by the VERIFY IP-MAC() module (i.e., ARP probe by the HIDS). The response for the ARP probe can be determined if $RSP_{IPD} == IP(HIDS)$ and $RSP_{MACD} == MAC_{HIDS}$ and $RSP_{IPS}$ has an entry in the Verification table. For such response packets, Algorithm 2 calls SPOOF-DETECROR() module.

If none of the above cases holds, the reply packet is then matched for a corresponding request in the Request-sent table, using its source IP $RSP_{IPS}$. If a match is found (i.e., $RSP_{IPS}== RQT_{IPD}[i]$), the $RSP_{IPS}$ is searched in the Authenticated bindings table. If a match is found and the corresponding MAC address in the table is same as $RSP_{MACS}$, the packet has genuine IP-MAC pair (which is already recorded in the Authenticated bindings table). In case of a mismatch in the MAC address (i.e., $RSP_{MACS} =6 AUTHT_{MACS}[ j]$) the packet may be spoofed with a wrong MAC address and hence the status flag is set as spoofed. If the $RSP_{IPS}$ is not present in the Authenticated bindings table, then an ARP probe is sent for verification by the VERIFY IP-MAC() module. If there was no corresponding request for the response packet in the Request-sent table, then it is an unsolicited response packet. Hence, the UNSOLICITED-RESPONSE-HANDLER() is called with the time at which such a response is received $\tau$.

**2.2.2 Algorithm 2: ARP RESPONSE HANDLER**

**Input:** *RSP* - ARP response packet, t - time at which *RSP* was received, Request-sent table, Verification table, authenticated bindings table

**Output:** Updated Response-received table, Status

1: **if** *RSP* is *malformed* **then**
2:    Status= *malformed*
3: **else**
4:    Add $RSP_{IPS}$, $RSP_{MACS}$ and $\tau$ to Response-received table
5:    **if** ($RSP_{IPS}$ == $RSP_{IPD}$) **then**
6:      Status= *Gratuitous*
7:      VERIFY IP-MAC(*RSP*, $\tau$)
8:    **else**
9:      **if** (($RSP_{IPD}$ == IP(HIDS) && $RSP_{MACD}$ == MAC(HIDS)) && ($RSP_{IPS}$ == $VRFT_{IPS}[k]$))(for some *k*, 1≤ *k*≤ $VRFT_{MAX}$ )) **then**
10:        EXIT
11:      **else**
12:        **if** ($RSP_{IPS}$ == $RQT_{IPD}[i]$ (for some *i*, $1 \leq i \leq RQ_{MAX}$ )) **then**
13:          **if** ($RSP_{IPS}$ == $AUTHT_{IPS}[j]$ (for some *j*, $1 \leq j \leq AUTHT_{MAX}$ )) **then**
14:            **if** ($RSP_{MACS}$ == $AUTHT_{MACS}[j]$) **then**
15:              Status=*Genuine*
16:            Else
17:              Status=*Spoofed*
18:            end if
19:          else
20:            VERIFY IP-MAC(*RSP*, $\tau$)
21:          end if
22:        else
23:          UNSOLICITED-RESPONSE-HANDLER( $\tau$)
24:        end if
25:      end if
26:    end if
27: **end if**

The main modules discussed in Algorithms1 and Algorithm 2 are assisted by three sub-modules namely, VERIFY IP-MAC(), SPOOF-DETECTOR() and UNSOLICITED-RESPONSE-HANDLER(). Now, we discuss these sub-modules in detail.

VERIFY IP-MAC() (Algorithm 3) sends ARP probes to verify the correctness of the IP-MAC pair given in the source of the request packet *RQP* or response packet *RSP*. Every time a probe is sent, its record is inserted in Verification table. Before, sending the ARP probe request, we need to verify if there is already such a request made by the HIDS and response is awaited. This can be verified by checking IP and MAC in the Verification table; if a match pair is found the module is exited. A spoofing may be attempted if IP matches the entry in the Verification table but MAC does not; in this case, the status is set as spoofed. This checking in the Verification table (before sending probe) limits the number of ARP probes to be sent for any known falsified IP-MAC address, thereby lowering extra ARP traffic. If the corresponding IP address is not found in the Verification table, a probe request is sent and the algorithm adds the IP and the MAC into the Verification table. At the same time SPOOF-DETECTOR() module is called which waits for a round trip time and then analyzes all the entries in the Response-received table collected within this period (for analyzing the probe responses).

### 2.2.3 Algorithm 3: VERIFY IP-MAC

**Input:** *RP*- ARP request/reply packet, t - time of arrival of *RSP*, Verification table **Output:** Updated Verification table, Status

 1: **if** ($RP_{IPS}\ 2\ VRFT_{IPS}[i]$) (for some $i$, $0 \leq i \leq VRFT_{MAX}$) **then**
 2:   **if** ($RP_{MACS} == VRFT_{MACS}[i]$) **then**
 3:      EXIT
 4:   **else**
 5:      Status=*Spoofed*
 6:   **end if**
 7: **else**
 8:   Send ARP *Probe Request* to $RP_{IPS}$
 9:   Add $RP_{IPS}$ and $RP_{MACS}$ to the Verification table
10:   SPOOF-DETECTOR(*RP*, τ)
11: **end if**

SPOOF-DETECTOR() (Algorithm 4) is called from VERIFY IP-MAC() after sending the ARP *Probe Request* to source IP of the packet to be checked for spoofing ($RP_{IPS}$). As discussed, it is assumed that all replies to the ARP probe will be sent within $T_{req}$ time. So, SPOOF-DETECTOR() waits for $T_{req}$ interval of time, thereby collecting all probe responses in the Response-received table. As it is assumed that non-comprised hosts will always respond to a probe, at least one response to the probe will arrive. In other words, in one of the replies to the probe, genuine MAC for the IP $RP_{IPS}$ would be present. Following that, Response-received table will be searched to find IP-MAC (source) pairs having IP of $RP_{IPS}$. If all IP-MAC pairs searched have same MAC, packet under question is not spoofed. In case of the packet being spoofed, more than one re-ply will arrive for the probe, one with genuine MAC and the other with spoofed MAC. The reason for assuming more than one reply in case of spoofing is explained as follows. Let a packet be spoofed as IP(of B)-MAC(of D). Now for the ARP probe to B, B will reply with IP(of B)-MAC(of B) leading to tracking the attacker (MAC (of D)). To avoid self identification, attacker D has to reply to all queries asking for B with spoofed IP-MAC pair IP(B)-MAC(D). The IDS has no clue whether IP(B)-MAC(D) or IP(B)-MAC(D) is genuine; only possibility of spoofing is detected.

If a spoofing attempt is determined the status is returned as spoofed and it is exited. If the packet is found genuine, Authenticated bindings table is updated with its source IP ($RP_{IPS}$) and the corresponding MAC.

### 2.2.4 Algorithm 4: SPOOF-DETECTOR

**Input:** *RP*- ARP request/reply packet, $T_{req}$ - Time required for arrival of all responses to an ARP probe (ARP request-reply round trip time), Response-received table **Output:** Updated Authenticated bindings table, Status

 1: Wait for $T_{req}$ time interval
 2: **if** ($RP_{IPS} == RST_{IPS}[i]$) && ($RP_{MACS} \neq RST_{MACS}[i]$)(for some $i, 1 \leq i \leq RST_{MAX}$) **then**
 3:   Status=*Spoofed*
 4:   EXIT
 5: **end if**
 6: Update Authenticated bindings table with $RP_{IPS},RP_{MACS}$

UNSOLICITED-RESPONSE-HANDLER() (Algorithm 5) is invoked whenever an unsolicited ARP reply packet is received (i.e., ARP reply packet did not find a matching ARP request in the Request-sent table) and is used for detection of denial of service (DoS) attacks. In general, ARP replies are received corresponding to the ARP requests. If more than a certain number of unsolicited responses are are sent to a host within a time window, it implies an attempt of DoS attack on the given host. Algorithm 5 maintains an "Unsolicited response counter (denoted as $URSP_{counter}$) for storing the number of unsolicited responses received by the host within a specified time interval ($\delta$) and declares DoS attack if the number of unsolicited ARP replies within a time interval ($\delta$) exceeds a preset threshold $DoS_{Th}$.

### 2.2.5 Algorithm 5: UNSOLICITED-RESPONSE-HANDLER

**Input:** $\tau$ - Time when *RSP* is received, $\delta$- Time window, $DoS_{Th}$- DoS Threshold, Un-solicited response table

**Output:** Status

```
 1: if (τ −URSPτ< δ ) then
 2:     URSPcounter++
 3:     URSPτ=τ
 4:     if (URSPTcounter  > DoSTh) then
 5:        Status=DoS
 6:        EXIT
 7:     end if
 8: else
 9:     URSPcounter=1
10:     URSPτ=τ
11: end if
```

## 2.3 An Example

In this sub-section we illustrate ARP response verification in normal and spoofed cases. Consider a network with four hosts - A, B, C, D and host D is the attacker. HIDS be installed on all the hosts which need to be secure.

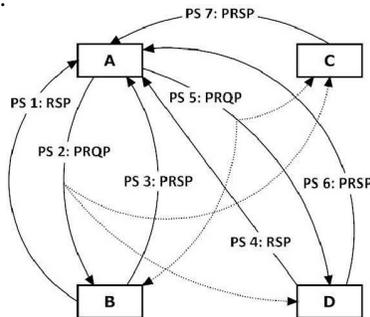

Fig. 1. Example of a Normal and Spoofed Reply

Figure 1 shows the sequence of packets (indicated with packet sequence numbers) injected in the LAN when (i) host B is sending a genuine reply to B with IP(B)-MAC(B) followed by ARP probe based verification (of the reply) by the HIDS at host A, (ii) attacker D is sending a spoofed reply as "IP(C)-MAC(D) " to host A and its verification by the HIDS at host A. The sequences of packets as recorded in Request-sent table, Response-received table, Verification table and Authenticated bindings table in the HIDS running in host A are shown in Table 1 - Table 4.

### 2.3.1 Genuine reply from B to A and its verification

– Packet Sequence (PS) 1: Reply is sent by B to A for informing its MAC address (to B). Assume this reply packet to be a gratuitous reply or a response to a request sent by A, so that it is not considered unsolicited. Response-received table is updated with a new entry IP(B)-MAC(B) .

International Journal of Network Security & Its Applications (IJNSA), Vol.3, No.3, May 2011- Packet Sequence 2: Since there is no entry for IP-MAC of B in Authenticated bindings table, the HIDS A will send an ARP Probe to verify its correctness and hence an entry is made in the Verification table.
- Packet Sequence 3: Following PS 2, SPOOF-DETECTOR() starts. Within $T_{req}$ only B will respond to this ARP Probe request and Authenticated bindings table is up-dated with the valid entry of IP-MAC of B.

### 2.3.2 Spoofed reply from D to A and its verification

- Packet Sequence 4: Let D respond to A with IP of C and its own MAC (D), which is recorded in the Response table. As in the above case, we consider this reply packet to be solicited.
- Packet Sequence 5: Since there is no entry for IP-MAC of C in Authenticated bindings table, the HIDS A will send an ARP probe to know C's MAC. Hence, IP(C)-MAC(D) is entered in the Verification table.
- Packet Sequence 6 and 7: SPOOF-DETECTOR() is executed. Within $T_{req}$, both C and attacker D will respond to the ARP Probe request (sent to know MAC of B) with their own MACs. These responses are recorded in the Response table.

  There are two entries in Response table for IP(C), one is MAC of C and the other is MAC of D. So response spoofing is detected by the HIDS running at A.

**Table 1.** Request-sent table

| PS | DST IP |
|----|--------|
| -  | -      |

**Table 2.** Response-received table

| PS | SRC IP | SRC MAC |
|----|--------|---------|
| 1  | IP B   | MAC B   |
| 3  | IP B   | MAC B   |
| 4  | IP C   | MAC D   |
| 6  | IP C   | MAC D   |
| 7  | IP C   | MAC C   |

**Table 3.** Verification table

| PS | IP   | MAC   |
|----|------|-------|
| 2  | IP B | MAC B |
| 5  | IP C | MAC D |

**Table 4.** Authenticated bindings table

| PS   | MAC   |
|------|-------|
| IP B | MAC B |

## 3. EXPERIMENTATION

The test bed created for our experiments consists of 5 machines running different operating systems. We name the machines with alphabets ranging from A-D. Machines A-D are running the following OSs: Ubuntu 9.04, Windows XP, Windows 2000 and Backtrack 4, respectively. The machine D with Backtrack 4 is acting as the attacker machine. HIDS is installed in all genuine machines.

The tables mentioned above are created in mysql database. The algorithms are implemented using C language. The HIDS has two major modules namely, Packet grabber and Packet injector. Packet grabber sniffs the packets from the host's network interface card (NIC), filters the ARP packets and invokes either the Algorithm 1 or Algorithm 2 depending upon the packet type - request or response. The Packet injector module generates the ARP probes necessary for the verification of IP-MAC pairs. Attack generation tools Ettercap, Cain and Abel were deployed in machine D and several scenarios of spoofing MAC addresses were attempted.

Table 5. Comparison of ARP Attack Detection Mechanisms

| ATTACKS | PROPOSED | ACTIVE [13] | X-ARP [12] | ARPWATCH [10] | ARPGUARD [11] |
|---|---|---|---|---|---|
| ARP spoofing | Y | Y | Y | Y | Y |
| ARP DoS | Y | N | N | N | N |
| Malformed Packets | Y | Y | Y | N | N |

In our experiments we tested our proposed scheme with several variants of LAN attack scenarios (including the one discussed in the example above). Table 5 presents the types of LAN attacks generated and detected successfully by the proposed scheme. Also, in the table we report the capabilities of other LAN attack detecting tools for these attacks.

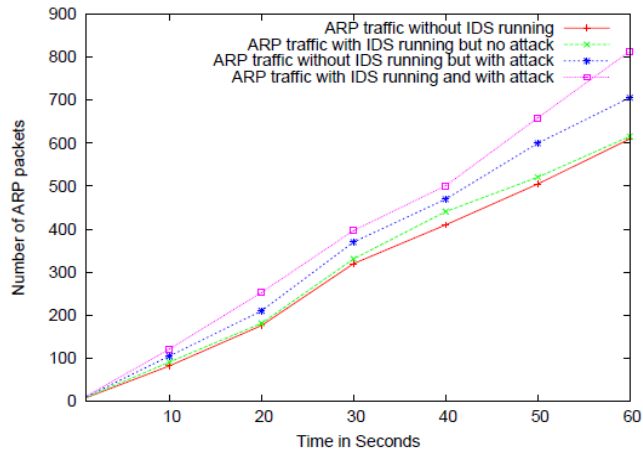

Fig. 2. ARP Traffic

Figure 2 shows the amount of ARP traffic generated in the experimentation in 4 cases. The first case is of normal operation in the absence of the IDS. Second case is when the IDS is running and there are no attacks generated in the network. Third case is when we injected 100 spoofed IP-MAC pairs into the LAN and IDS is not running. Fourth case is when we injected 100 spoofed IP-MAC pairs into the LAN with IDS running. We notice almost same amount of ARP traffic under normal situation with and without IDS running. Once genuine IP-MAC pairs are identified (by probing) they are stored in Authenticated bindings table. Following that no probes are required to be sent for any ARP request/reply from these IP-MAC pairs. In case of attack, a little extra traffic is generated by our IDS for the probes. With each spoofed ARP packet, our IDS sends a probe request and expects at least two replies (one from normal and the other from the attacker), there by adding only three ARP packets for each spoofed packet.

## 4. ANALYSIS OF COMPLETENESS AND CORRECTNESS

In this section we prove the completeness of the algorithm IDS using different scenarios of ARP spoofing attacks. In this paper, hosts are identified by alphabets, *a* for example, and IP (MAC) address of the host is represented by IP(*a*) (MAC(*a*)). Before the proof, certain definitions are given.

### 4.1 ARP spoofing attack

In ARP spoofing attack a malicious host *m* sends an ARP request/response packet to another host *p* in the LAN with falsified IP-MAC pair. In the response/request packet being sent, IP

172

address of host $v$ IP($v$) is associated with MAC address of host $k$ MAC($k$), where $v \neq k$ and $v;k \neq p$. In other words, an ARP request or response packet is created and sent by $m$ which has source IP-MAC pair as IP($v$)-MAC($k$). When response/request packet with IP($v$)-MAC($k$) is sent to $p$, it updates its cache with IP($v$)-MAC($k$) and all packets $p$ wants to send $v$ will reach $k$. The attack is said to be created with falsified IP-MAC pair "IP($v$)-MAC($k$)" against victim $v$ by $m$.

## 4.2 Victim Machine

A machine in the LAN whose network traffic can be redirected to some other machine is called the victim machine. In Definition 4, host $v$ is the victim machine as traffic being sent by $p$ to $v$ is redirected to $k$.

An interesting case occurs if ARP attack is created with IP($v$)-MAC($k$), where $v$ is the malicious host (i.e., $v = m$). In this case, malicious host $m$ sends an ARP request/response packet with IP($m$)-MAC($k$) ($m \neq k$) to another host $p$. So all packets $p$ wants to send $m$ will reach $k$. So in this case attacker becomes the victim.

We assume the following in our arguments.

– The set of IP addresses in the LAN corresponding to which the hosts are up is $I$. – The set of IP addresses in the LAN corresponding to which the hosts are down is $I'$.

We will study completeness of the IDS in two scenarios of the LAN. In the first case let all the machines in the LAN are up and running. In the second case, there may be some machines which are down. In this case, we will also see the situation when such machines are powered up after an attack is launched against them.

### 4.2.1 Theorem 1

*The IDS detects all ARP spoofing attacks except the case where attacker becomes the victim, when all IP address in I are used.*

*Proof.* We prove the theorem by enumerating all possible combinations of IP-MAC pairs generated by malicious host. Let there be a single malicious host $m$ in the LAN having IP-MAC pair IP($m$)-MAC($m$); later we will show that the theorem holds for multiple malicious hosts also. It is assumed that Authenticated and Spoofed tables are empty.

There are 5 cases for IP-MAC combinations that can be generated by $m$–(A) IP($m$)-MAC($m$), (B) IP($v$)-MAC($v$) (IP($v$) $\in I \neq$ IP($m$)), (C) IP($v$)-MAC($m$) (IP($v$) $\in I \neq$ IP($m$)), (D) IP($m$)-MAC($v$) ($v \neq m$), (E) IP($v$)-MAC($k$) (IP($v$) $\in I$, $v \neq k$ and $v;k \neq m$).

Now these cases are analyzed one by one. (A), (B) do not correspond to spoofed cases. The rest are analyzed as follows. The analysis is shown when the malicious host uses a request packet for sending the falsified IP-MAC pair. Same argument will hold when the malicious host uses a response packet for sending the falsified IP-MAC pair.

#### 4.2.1.1 Case 1–IP($v$)-MAC($m$) (IP($v$) $\in I \neq$ IP($m$))

If ARP request packet with IP($v$)-MAC($m$) is sent by $m$ to the host, then REQUEST-HANDLER() generates a *RQP* event. As Authenticated and Spoofed tables are empty, probe request is sent by IDS for verification and *PRQP* event is generated. As the probe request is broadcast it is received by all hosts. So this probe request is also received by the malicious host $m$ and may respond in the ways enumerated below. However, in all cases the genuine host will reply to the probe using an ARP response packet having source IP-MAC pair as IP($v$)-MAC($v$); for this the RESPONSE-HANDLER() generates a *PRSP* event, whose MAC address is not same as that of MAC address of the request being verified.

**Variations of Responses of** $m$

- **(a)** *m* may not give any reply. So there will be only one reply having IP(*v*)-MAC(*v*) (from genuine host *v*). So, in all, the following sequence of events is received by the IDS – *RQP*,*PRQP*,*PRSP*; the MAC address of *PRSP* (MAC address of IP-MAC pair sent with probe response, MAC(*v*)) is not same as that of MAC address of the request packet being verified (having MAC address MAC(*m*)).

RQP: The RESPONSE-HANDLER() will check for gratuitous packet which will be found false as source IP and destination IP's are not same (Algorithm 2, Step 5). Then it will compare the IP(*v*)-MAC(*v*) pair with that of IDS (Algorithm 2, Step 9), but no match will be found. Then it will check request sent table for a match with IP(*v*) (Algorithm 2, Step 12); no match will be found. Next on comparison with authentication binding table (Algorithm 2, Step 13), also no match will be found (assuming RSP was not verified earlier). This will result in calling of VERIFY IP-MAC(RSP,$\tau$) (Algorithm 2, Step 20) which then will verify if IP(*v*) is in verification table (Algorithm 3, Step 1). Since it will not find any match, a probe will be sent (Algorithm 3, Step 8) to IP(*v*)and IP(*v*), MAC(*m*) will be recorded in verification table (Algorithm 3, Step 9) and call the SPOOF-DETECTOR(RP,$\tau$) (Algorithm 3, Step 10). The SPOOF-DETECTOR will wait for $T_{req}$ time interval (Algorithm 4, Step 1) and then verify IP(*v*) with IP of *PRSP* (Algorithm 4, Step 2), will find a match. Also it will compare the MAC(*v*) (Algorithm 4, Step 2), will find a mismatch for the corresponding MAC address, resulting into the declaration of the packet be spoofed with a wrong MAC address and hence the status flag will be set as spoofed. *So IP(v)-MAC(m) is correctly identified to be spoofed.*

Also, the reply from *v* with IP(*v*)-MAC(*v*) leads to tracking the attacker (MAC(*m*)). To avoid self identification, attacker *m* has to give a single reply to all queries asking for MAC of *v* with spoofed IP-MAC pair IP(*v*)-MAC(*m*); this mimics as if IP(*v*)-MAC(*m*) was normal. The IDS has no clue whether IP(*v*)-MAC(*v*) or IP(*v*)-MAC(*m*) is genuine; only possibility of spoofing is detected. In other words, to avoid being detected, if the attacker sends a spoofed packet IP(*v*)-MAC(*m*) say, then for all ARP requests for MAC of *IP(v)* it would send a reply with the same spoofed MAC address (i.e., MAC(*m*)) that it has used in spoofing. This behavior of attacker is assumed for all queries for MAC address it has spoofed. With this assumption, the case of one reply having IP(*v*)-MAC(*m*) from *m* is analyzed as follows.

- **(b)** One reply having IP(*v*)-MAC(*m*) from *m*. It may be noted that the sequence in which host *v* and host *m* respond to the probe request is not fixed. Let *v* respond before *m*. The two response packets (one from *v* and the other from *m*) are processed by RESPONSE-HANDLER() as two *PRSP* events. So, in all, the following sequence of events is received by the IDS–*RQP*,*PRQP*,*PRSP*,*PRSP*; the MAC address of first *PRSP* is different from the of MAC address of the request being verified and the MAC address of second *PRSP* is same.

- **(c)** The RESPONSE-HANDLER() will check for gratuitous packet which will be found false as source IP and destination IP's are not same. Then it will com-pare the IP(*v*)-MAC(*v*) pair with that of IDS, but no match will be found. Then it will check request sent table for a match with IP(*v*), will not find a match. Next on comparison with authentication binding table, also no match will be found (assuming it was not verified earlier). This will result in calling of VERIFY IP-MAC(RSP,$\tau$) which then will verify if IP(*v*) is in verification table. Since it will not find any match, a probe will be sent to IP(*v*)and record IP(*v*), MAC(*v*) in verification table and call the SPOOF-DETECTOR(RP,$\tau$). The SPOOF-DETECTOR will wait for $T_{req}$ time interval and then verify IP(*v*) with IP of *PRSP*, will find a match, compare the MAC(*v*), will find a mismatch for the corresponding MAC address, resulting into the declaration of the packet be Spoofed and hence the status flag will be set as Spoofed. *So IP(v)-MAC(m) is* *correctly identified to be spoofed.*

Now, let *m* respond before *v*. In this situation, sequence of events received is by the IDS is– *RQP*,*PRQP*,*PRSP*,*PRSP*, where the MAC address of first *PRSP* is same as the MAC address of the request being verified and the MAC address of second *PRSP* is different.

The RESPONSE-HANDLER() will check for gratuitous packet which will be found false as source IP and destination IP's are not same. Then it will com-pare the IP(*v*)-MAC(*v*) pair with that of IDS, but no match will be found. Then it will check request sent table for a match with IP(*v*), will not find a match. Next on comparison with authentication binding table, also no match will be found (assuming it was not verified earlier). This will result in calling of VERIFY IP-MAC(RSP,$\tau$) which then will verify if IP(*v*) is in verification table. Since it will not find any match, a probe will be sent to IP(*v*)and record

IP(*v*), MAC(*v*) in verification table and call the SPOOF-DETECTOR(RP,$\tau$). The SPOOF-DETECTOR will wait for $T_{req}$ time interval and then verify IP(*v*) with IP of *PRSP*, will find a match, compare the MAC(*v*), will find a match for the corresponding MAC address, resulting into the declaration of the packet be Genuine but as soon as the second PRSP from *v* will arrive and the algorithm will execute in the same fashion but a mismatch will be detected in this step and Spoofing will be detected and hence the status flag will be set as Spoofed.

*So IP(v)-MAC(m) is correctly identified to be spoofed.*

### 4.2.1.2 Case 2–IP(*m*)-MAC(*v*) (*v* ≠ *m*)

REQUEST-HANDLER() generates a *RQP* event on receipt of the request packet sent by *m* having IP(*m*)-MAC(*v*). Following that probe request is sent by IDS for verification of MAC address (MAC(*v*)) associated with IP address (IP(*m*)) of the request packet and *PRQP* event is generated. This probe request sent to query MAC address of IP(*m*). Even of the request is received by all hosts, as all hosts except *m* are genuine, only *m* would respond. According to the assumption in Case-1, *m* will respond to the probe as IP(*m*)-MAC(*v*); the case is analyzed as follows.

(a) One reply with IP(*m*)-MAC(*v*). REQUEST-HANDLER() generates event *PRSP*.

So, in all, the following sequence of events is received by the IDS –*RQP*,*PRQP*,*PRSP*; the MAC address of *PRSP* is same as that of MAC address of the request being verified.

The RESPONSE-HANDLER() will check for gratuitous packet which will be found false as source IP and destination IP's are not same. Then it will com-pare the IP(*v*)-MAC(*v*) pair with that of IDS, but no match will be found. Then it will check request sent table for a match with IP(*v*), will not find a match. Next on comparison with authentication binding table, also no match will be found (assuming it was not verified earlier). This will result in calling of VERIFY IP-MAC(RSP,$\tau$) which then will verify if IP(*v*) is in verification table. Since it will not find any match, a probe will be sent to IP(*v*)and record IP(*v*), MAC(*v*) in verification table and call the SPOOF-DETECTOR(RP, $\tau$). The SPOOF-DETECTOR will wait for $T_{req}$ time interval and then verify IP(*v*) with IP of *PRSP*, will find a match, compare the MAC(*v*), will find a match for the corresponding MAC address, resulting into the declaration of the packet be Genuine.

*So IP(m)-MAC(v) is incorrectly identified to be genuine.* It may be noted that if response to the probe has same MAC address as that of the request being verified, it is determined to be genuine. In other words, if no response to the probe has different MAC address compared to the request being verified, it is determined to be genuine.

175

**4.2.1.3 Case 3–IP(*v*)-MAC(*k*) (*IP(v);IP(k)* $\in$, *v* $\neq$ *k* and *v;k* $\neq$ *m*)**

REQUEST-HANDLER() generates a *RQP* event on receipt of the request packet sent by *m* having IP(*v*)-MAC(*k*). Following that probe request is sent by IDS for verification and *PRQP* event is generated. As the probe request is broadcast it is received by all hosts. As *v* is genuine it will respond by a ARP response packet having IP(*v*)-MAC(*v*), whose MAC is different than the one in the request packet being verified. The attacker will respond to this probe as IP(*v*)-MAC(*k*). Attack is detected by the IDS because there is at least one repone to the *PRQP* whose MAC is different from the request packet being verified. *So IP (v)-MAC(k) is correctly identified to be spoofed.*

From the enumeration above, only Case 2 Subcase (a) is the condition when spoofing is detected as genuine; i.e., spoofing attack cannot be detected. Case 2 corresponds to IP(*m*)-MAC(*v*) (*v* $\neq$ *m*). Let *m* send a request with IP (*m*)-MAC(*v*) (*v* $\neq$ *m*) to host *p*, which updates its cache accordingly. So all traffic *p* wants to send *m* will reach *v*; so Case 2 corresponds to condition where *m* is also the victim (in addition to being a malicious host).

From the theorem the following corollary follows.

**Corollary 1**

*If all responses to the probe sent by IDS for verifying any request/response packet has same MAC address as that of the packet being verified, the IDS determines normal condition. If at least one response to the probe sent by IDS for verifying any request/response packet has different MAC address compared to the packet being verified, then IDS determines spoofed condition.*

*Proof.* Follows from construction of IDS and illustrated in Theorem 1.

In the next theorem we will show that Theorem 1 also holds when there is more than one malicious host. Before that Theorem 1 is restated as follows:

Let IP(*v*)-MAC(*k*) (*IP(v) 2 I =6 IP(k)*) be sent in a spoofed request/response by malicious host *m*. Spoofing can be detected if *v* $\neq$ *m*.

The elaborate proof given above for Theorem 1 can be summarized as follows.

As shown in Corollary 1, spoofing is detected if at least one response to the probe sent by IDS for verification has different MAC address compared to the packet being verified. If IP (*v*)-MAC(*k*) is sent (by *m*), then IDS sends a probe (by broadcast) to query the MAC address associated with IP(*v*). If *v* $\neq$ *m*, then *v* is genuine and as *IP(v) 2 I* (i.e., *v* is up) it will reply with IP(*v*)-MAC(*v*); as the reply has different MAC address compared to the packet being verified, spoofing can be detected.

Hoverer, if *v = m*, then *v* is itself the attacker. The IDS probe request is sent to query MAC address of IP(*v = m*) and no genuine host would reply. The malicious host *m* will reply with spoofed ARP reply whose MAC address is deliberately kept same as the one in the packet being verified. So IP(*v*)-MAC(*k*) is falsely determined to be genuine. It may be noted that IP(*v = m*)-MAC(*k*) is the case where attacker is the victim.

**4.2.2 Theorem 2**

*Let the set of malicious hosts in a LAN be M. Let IP(v)-MAC(k) (IP(v)* $\in$ *I* $\neq$ *IP(k)) be sent in a spoofed request/response by malicious host m 2 M. Spoofing can be detected if v 2= M.*

*Proof.* As shown in Corollary 1, spoofing is detected if at least one response to the probe sent by IDS for verification has different MAC address compared to the packet being verified. If IP ($v$) - MAC ($k$) is sent (by $m$), then IDS sends a probe to query MAC.

Address associated with IP($v$). As $v \notin M$, IP($v$) $\in$ (i.e., $v$ is up) and malicious hosts cannot stop $v$ from responding, it will reply with IP($v$)-MAC($v$). Along with the reply from $v$, other malicious hosts can also reply to the probe request, however, cannot stop detection of spoofing because reply sent by $v$ has different (correct) MAC address com-pared to the packet being verified (which has spoofed MAC address).

Next we will study completeness of the IDS in the second scenario where some IPs in LAN may not be used, i.e., some of the machines are down. In this case we will also see the situation when such machines are powered up.

### 4.2.3 Theorem 3

*The IDS detects all ARP spoofing attacks except the cases (i) where attacker becomes the victim (i.e., IP(m)-MAC(v), $v \neq m$), and (ii) the IP address used in the spoofing packet corresponds to a machine which is down (i.e., IP(v)-MAC(k), IP(v) $\in I' \neq IP(k)$)*

*Proof.* We prove the theorem by enumerating all possible combinations of IP-MAC pairs generated by the malicious host. Let there be a single malicious host $m$ in the LAN having IP-MAC pair IP($m$)-MAC($m$).[2] It is assumed that Authenticated and Spoofed tables are empty.

There are 8 cases for IP-MAC combinations that can be generated by $m$–(A) IP($m$)-MAC($m$), (B) IP($v$)-MAC($v$) (IP($v$) $\in I' \neq$ IP($m$)), (C) IP($v$)-MAC($v$) (IP($v$) $\in I' \neq$ IP($m$)), (D) IP($v$)-MAC($m$) (IP($v$) $\in I' \neq$ IP($m$)), (E) IP($v$)-MAC($m$) (IP($v$) $\in I' \neq$ IP($m$)), (F) IP($m$)-MAC($v$) ($v \neq m$), (G) IP($v$)-MAC($k$) (IP($v$) $\in I'$, $v \neq k$ and $v;k \neq m$), (H) IP($v$)-MAC($k$) (IP($v$) $\in I'$, $v \neq k$ and $v;k \neq m$).

Case (A), (B), (C) do not correspond to spoofed cases. Case (D),(G) involve IP address IP($v$) $\neq I$; so this can be handled similarly as in Theorem 1. Also, Case (F) can be handled similarly as in Theorem 1 because it involves IP address $m \in I$ (attacker is the victim).

The rest of the cases ((E),(H)) are analyzed as follows. As in Theorem 1 the analysis is shown when the malicious host uses a request packet for sending the falsified IP-MAC pair. Same argument will hold when the malicious host uses a response packet for sending the falsified IP-MAC pair.

#### 4.2.3.1 Case 1–IP($v$)-MAC($m$) (IP($v$) $\in I' \neq IP(m)$)

REQUEST-HANDLER() generates a *RQP* event on receiving the packet with IP($v$)-MAC($m$). A probe request is sent by IDS for verification and *PRQP* event is generated. As machine with IP($v$) is not up, there would not be any reply from $v$ with IP($v$)-MAC($v$) (that has different MAC address than the packet being verified). Also, the malicious host $m$ will send a reply with same MAC address. So, there is a condition when only one response to the probe request is received that has same MAC address as that of the packet being verified. By Corollary 1, *IP(v)-MAC(m) is incorrectly identified to be genuine*.

#### 4.2.3.2 Case 2–IP($v$)-MAC($k$) (IP($v$) $\in I'$, $v \neq k$ and $v;k \neq m$)

This situation is similar to Case-1 (of Theorem 3) above. *IP (v)-MAC(k) is incorrectly identified to be genuine*.

In Theorem 3 we have seen two conditions when a spoofed packet is determined to be genuine. IP address in such spoofed packets corresponds to system(s) which are down. Now we will see the condition when such a system comes up.

Let a spoofed packet having IP($v$)-MAC($k$) (IP($v$)$\in I' \neq IP(k)$) be detected as genuine; so IP($v$)-

MAC(*k*) is entered in Authenticated Table. After, *v* comes up, it sends a gratuitous request with IP(*v*)-MAC(*v*); REQUEST-HANDLER() generates a *RQP* event. Following that a probe request is sent to verify to the gratuitous request thereby generating event *PRQP*. Host *v* responds to the probe request with IP(*v*)-MAC(*v*). Now malicious host will respond to the probe by IP(*v*)-MAC(*k*) ($k \neq v$), which has different MAC address corresponding to the gratuitous request being verified. So, gratuitous request is incorrectly determined to be spoofed and the falsified IP-MAC pair IP(*v*)-MAC(*k*) ($v \neq k$) remains to be kept in the Authenticated table.

The following points can be deduced from Theorem 1 and Theorem 3 regarding the consequences of the cases when a spoofed request/reply is determined to be genuine (i.e., false negative cases).

**Theorem 1**

Spoofed IP-MAC pairs for the case "when attacker itself is the victim", is missed to be detected. This does not lead to a serious consequence because no genuine host can be victimized (by diverting its traffic to some other host). So other attacks like man-in-the-middle, denial of service etc. which require diverting traffic sent to genuine hosts (to malicious hosts) cannot be launched.

**Theorem 3**

Spoofed IP-MAC pairs for cases "(i) when attacker itself is the victim" and "(ii)IP address used in a spoofed packet corresponds to a machine which is down" are missed to be detected. Further, even after the machine comes up, spoofing cannot be detected. Case (ii) may lead to serious consequence as traffic intended to a genuine host would be diverted to malicious host.

## 5. CONCLUSIONS

In this paper, we have presented an HIDS for detecting some of the LAN specific at-tacks and verified the same under all possible circumstances. The scheme uses an active probing mechanism and does not violate the principles of network layering architecture. This being a software based approach does not require any additional hardware to operate.

At present the scheme can only detect the attacks. In other words, in case of spoofing it can only determine the conflicting IP-MAC pairs without differentiating the spoofed IP-MAC and genuine IP-MAC pair. If to some extent diagnosis capability can be pro-vided in the scheme, some remedial action against the attacker can be taken.

## 5. ACKNOWLEDGEMENT

**Authors**

**Ferdous A Barbhuiya** is a PhD student at Indian Institute of Technology Guwahati doing research in the field of "Intrusion Detection System" He received the B.E. Degree in 2001 from Jorhat Engineering College, Jorhat, Assam, India, and the M.Tech. degree in 2007 from the Indian Institute of Technology Guwahati, India, and currently a Project Fellow in Indian Institute of Technology Guwahati. He has worked as a scientific office in IIT Guwahati during the period of June 2002 to Dec 2007. He has also worked in GlobalLogic Inc from Dec 2007 to July 2009 and also contributed in the conceptualization of products by many startup companies. His current research interests include system and network security, failure diagnosis and discrete event systems, sensor fusion etc.

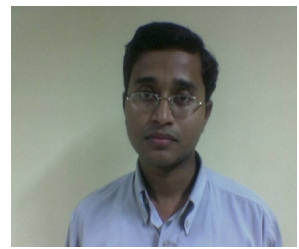

**Santosh Biswas** received B.E degree from National Institute of Technology, Durgapur, in 2001. He has completed his MS from the department of electrical engineering, Indian Institute of Technology Kharagpur with highest institute CGPA in the year 2004. He has completed PhD from the department of computer science and engineering of the same institute in the year 2008. At present he is an assistant professor in the department of computer science and engineering, IIT Guwahati. His research interests include networking, VLSI testing and design for testability, discrete-event systems and embedded systems. He has published about 50 research papers. He is the recipient of Infineon India best master's thesis award, 2005–2006. His biography has been published in Marquis Who's Who in science and engineering, 2006–2007.

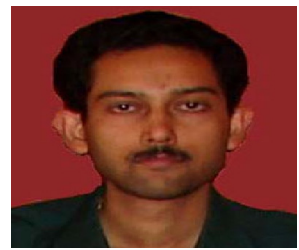

**Sukumar Nandi** received BSc (Physics), BTech and MTech from Calcutta University in 1984, 1987 and 1989 respectively. He received the PhD degree in Computer Science and Engineering from Indian Institute of Technology Kharagpur in 1995. In 1989–1990 he was a faculty in Birla Institute of Technology, Mesra, Ranchi, India. During 1991–1995, he was a scientific officer in Computer Science and Engineering, Indian Institute of Technology Kharagpur. In 1995 he joined Indian Institute of Technology Guwahati as an Assistant Professor in Computer Science and Engineering. Subsequently, he became Associate Professor in 1998 and Professor in 2002. He was in School of Computer Engineering, Nanyang Technological University, Singapore as Visiting Senior Fellow for one year (2002–2003). He was member of Board of Governor, Indian Institute of Technology Guwahati for 2005 and 2006. He was General Vice-Chair of 8th International Conference on Distributed Computing and Networking 2006. He was General Co-Chair of the 15th International Conference on Advance Computing and Communication 2007. He is also involved in several international conferences as member of advisory board/ Technical Programme Committee. He is reviewer of several international journals and conferences. He is co-author of a book titled "Theory and Application of Cellular Automata" published by IEEE Computer Society. He has published more than 150 Journals/Conferences papers. His research interests are Computer Networks (Traffic Engineering, Wireless Networks), Computer and Network security and Data mining. He is Senior Member of IEEE and Fellow of the Institution of Engineers (India)

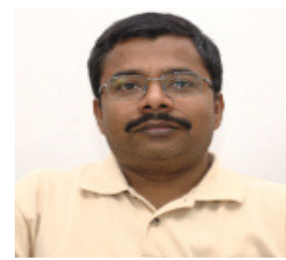